\documentclass{elsart1p}

\usepackage{graphicx}

\usepackage{amssymb}

\begin{document}

\begin{frontmatter}



\title{Measurement of electro-magnetic radiation at RHIC-PHENIX} 


\author{Takao Sakaguchi}
\ead{takao@bnl.gov}
\address{Physics Department, Brookhaven National Laboratory, Upton, NY 11973-5000, U.S.A.}

\author{for the PHENIX Collaboration}

\begin{abstract}
Recent results on direct photons and dileptons from the PHENIX experiment
opened up a possibility of landscaping electro-magnetic radiation
over various kinetic energies in heavy ion collisions.
Results on direct photon measurement in Au+Au collisions at
$\sqrt{s_{NN}}$=200\,GeV are discussed from the point of view of
structure function and isospin effect.
The first measurement of
direct photons at $\sqrt{s_{NN}}$=62.4\,GeV in the same collisional
system suggested that these effects are existing and would
manifest at $p_T>$16\,GeV/$c$ at $\sqrt{s_{NN}}$=200\,GeV.
\end{abstract}

\end{frontmatter}

%
\section{Introduction}\label{intro}
\begin{figure}[htb]
\centering
\begin{minipage}{58mm}
\includegraphics*[angle=270,width=5.0cm]{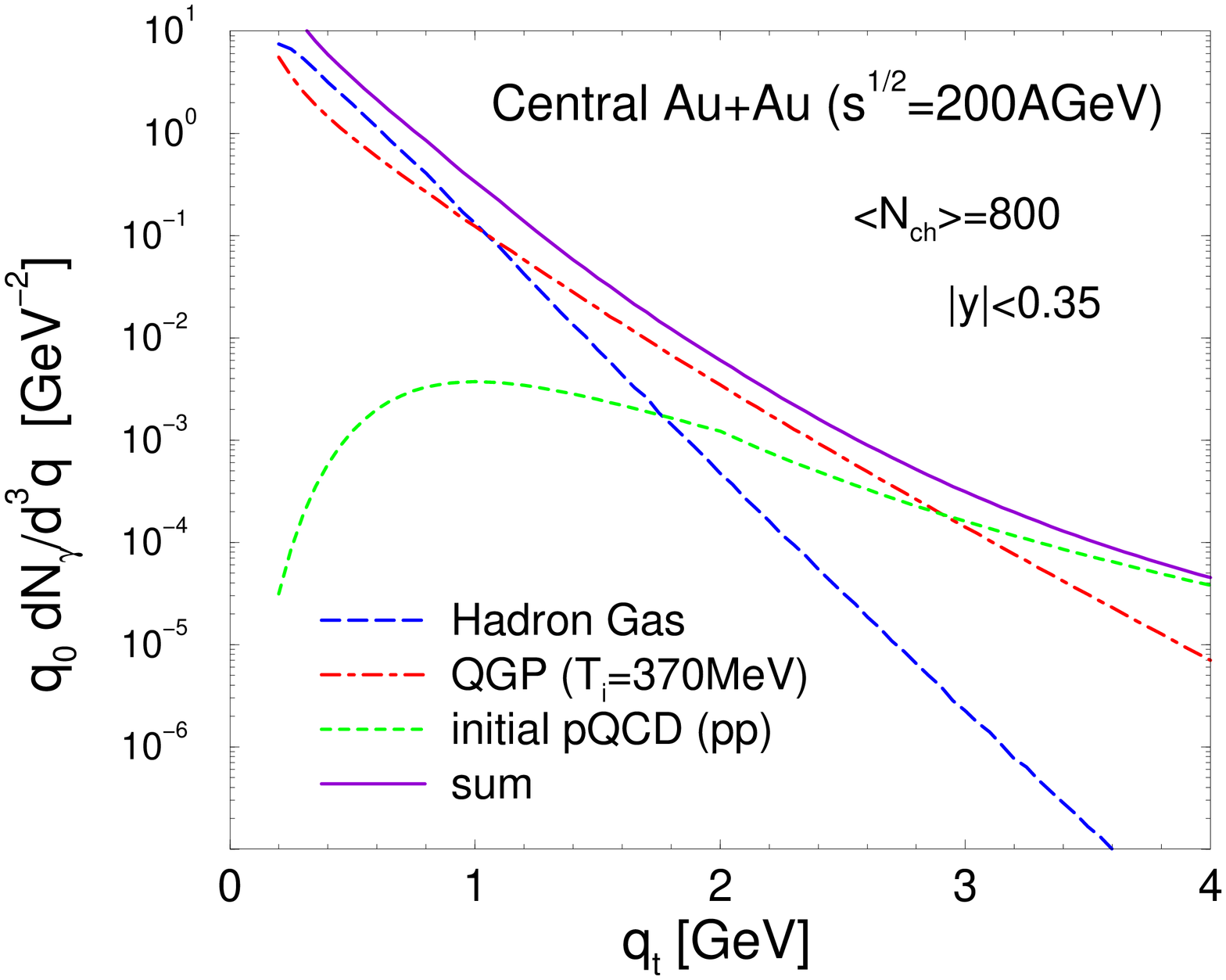}
\end{minipage}
\hspace{5mm}
\begin{minipage}{65mm}
\includegraphics*[width=5.5cm]{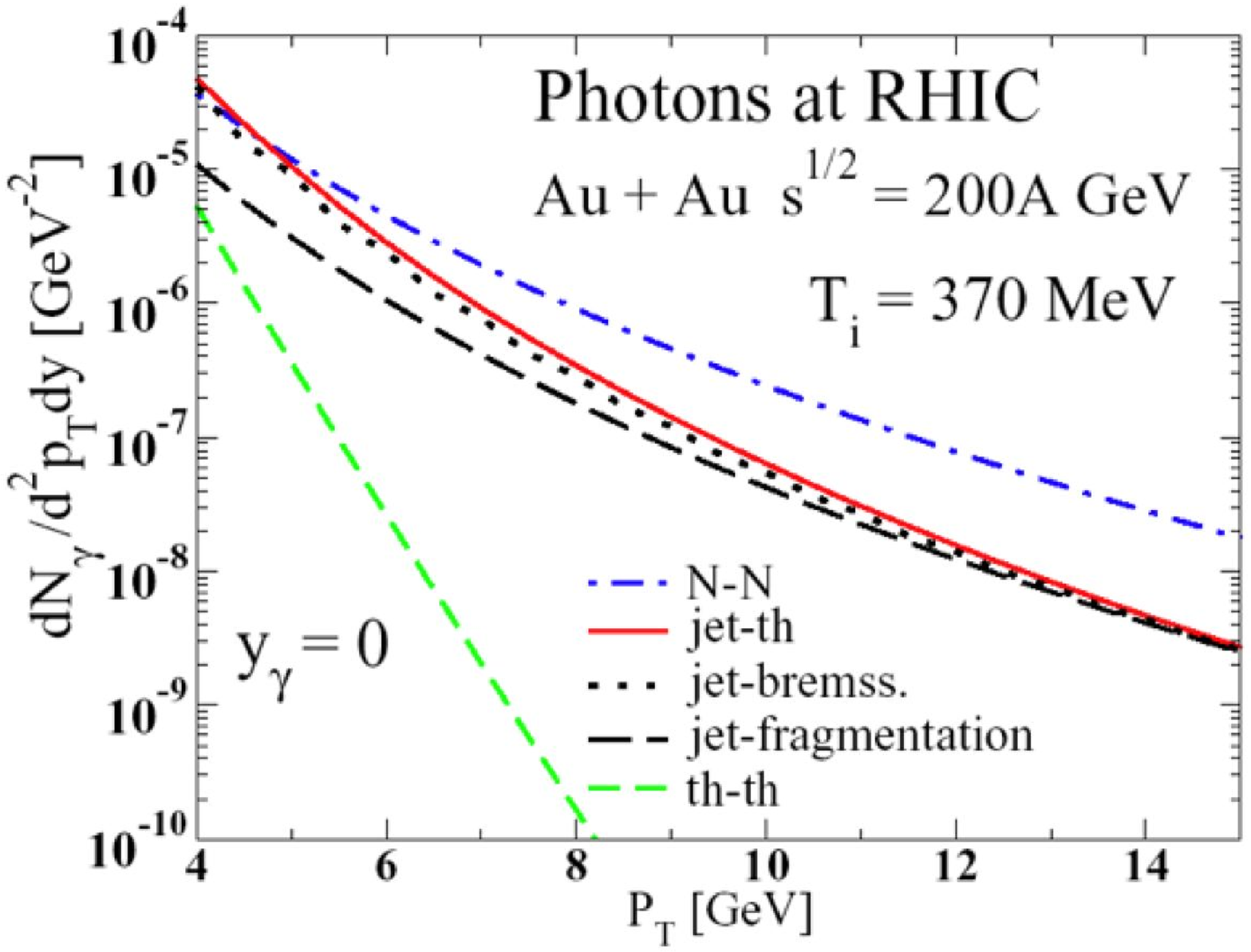}
\end{minipage}
\vspace{-2mm}
\caption[]{Predictions on direct photon production (a) at low $p_T$ (left) and (b) at high $p_T$ in Au+Au collisions at $\sqrt{s}$=200\,GeV (right).}
\label{figphotonPred}
\end{figure}

Electro-magnetic radiation is an excellent probe for extracting
thermodynamical information of a matter produced in nucleus-nucleus
collisions~\cite{Stankus:2005eq}. They are emitted from all the stages of
collisions, and don't interact strongly with medium once
produced. As stated in several literatures~\cite{Stankus:2005eq,ref1,ref2},
the electro-magnetic radiations stand for a thermal radiation from
the matter or a prompt emission from the initial stage. They are primarily
produced through a Compton scattering of quarks and gluons
($qg\rightarrow q \gamma$) or an annihilation of quarks and anti-quarks
($q\overline{q} \rightarrow g \gamma$) as leading order processes,
and the next leading order (NLO)
process is dominated by bremsstrahlung (fragment) ($qg \rightarrow qg\gamma$).
A calculation predicts that the photon contribution from a quark gluon plasma
(QGP) state is predominant in the transverse momentum ($p_T$) range of
1$<p_T<$3\,GeV/$c$~\cite{ref3}. For $p_T<$1\,GeV/$c$, the signal from hadron
rescattering process is dominant.
There is also a prediction of a jet-photon conversion process, which occurs,
if QGP is formed, by a secondary interaction of a hard scattered parton with
thermal partons in the medium~\cite{ref4}. The predictions are shown
in Fig.~\ref{figphotonPred}.

One of the big successes by now is the observation of high $p_T$ direct
photons produced in the initial hard scattering~\cite{ref2}.
The high $p_T$ hadron suppression found at RHIC is interpreted as
a consequence of an energy loss of hard-scattered partons
in the hot and dense medium. It was strongly supported by the fact that
the high $p_T$ direct photons are not suppressed and well described by
a NLO pQCD calculation~\cite{ref5}.

In this paper, the recent measurement on direct photons is focused, out of
various electro-magnetic radiation measurements from the PHENIX
experiment. The recent dilepton measurement at RHIC has just been
published~\cite{ref6}. A discussion on comprehensive understanding of
electro-magnetic radiations is presented in~\cite{ref7}.

\section{Direct photon measurement in Au+Au collisions at $\sqrt{s_{NN}}$=200\,GeV}
\begin{figure}[htb]
\centering
\begin{minipage}{65mm}
\includegraphics*[width=6.4cm]{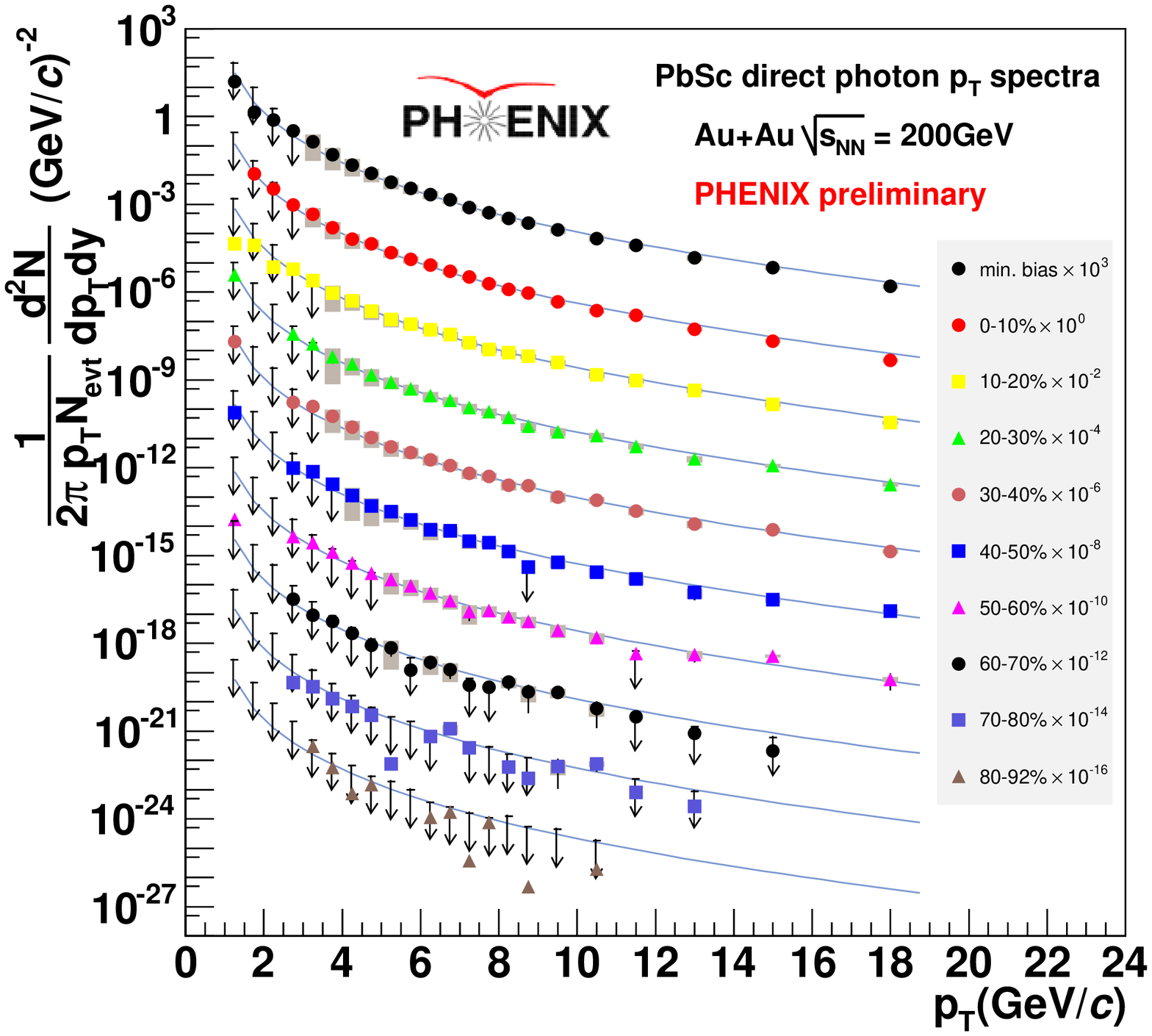}
\end{minipage}
\begin{minipage}{65mm}
\includegraphics*[width=6.2cm]{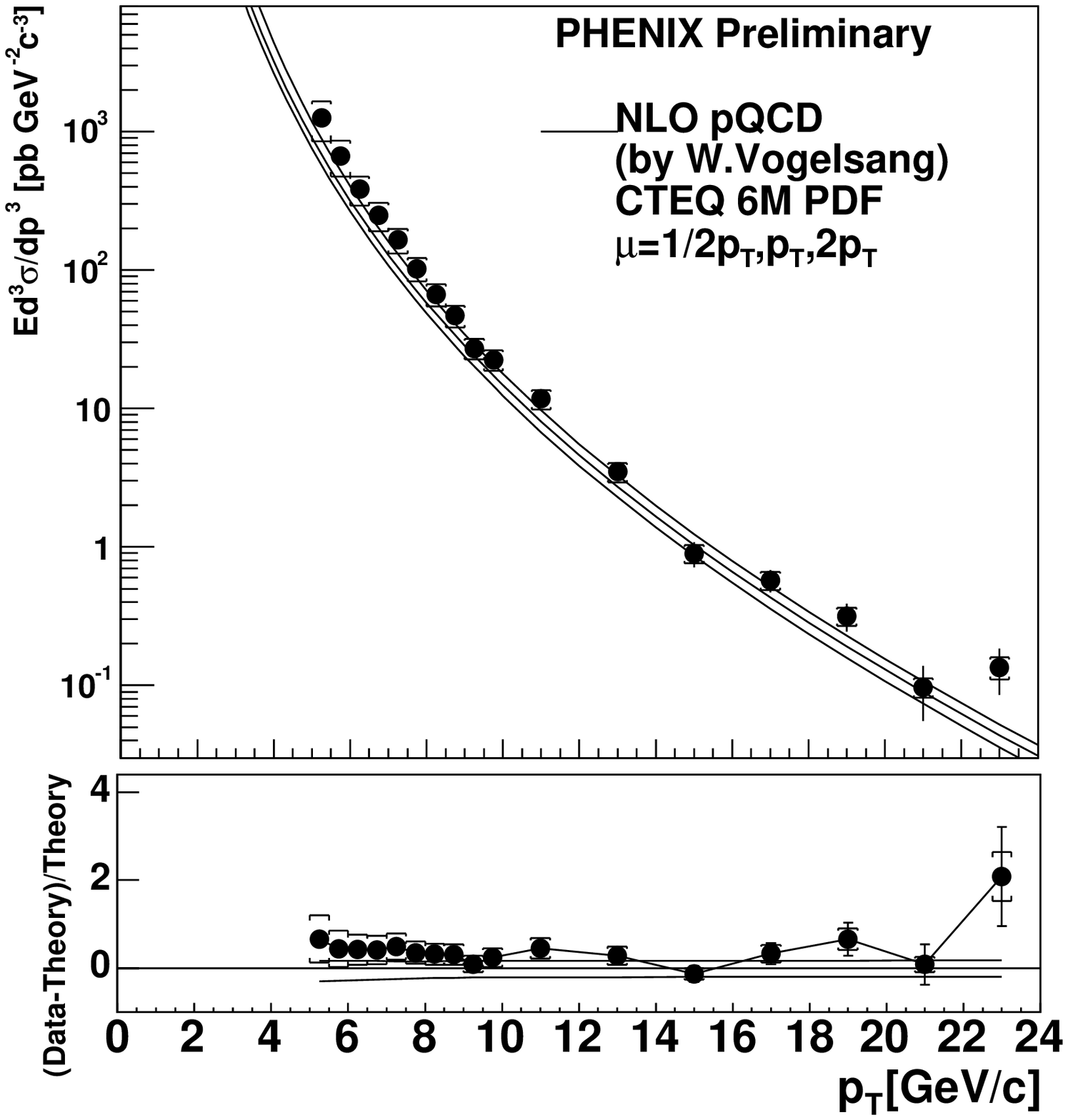}
\end{minipage}
\vspace{-2mm}
\caption[]{(a) Direct photon $p_T$ spectra in Au+Au collisions at $\sqrt{s_{NN}}$=200\,GeV (left) and (b) in p+p collisions at $\sqrt{s}$=200\,GeV (right).}
\label{figphotonspectra}
\end{figure}

\begin{figure}[htb]
\centering
\begin{minipage}{60mm}
\includegraphics*[width=6.0cm]{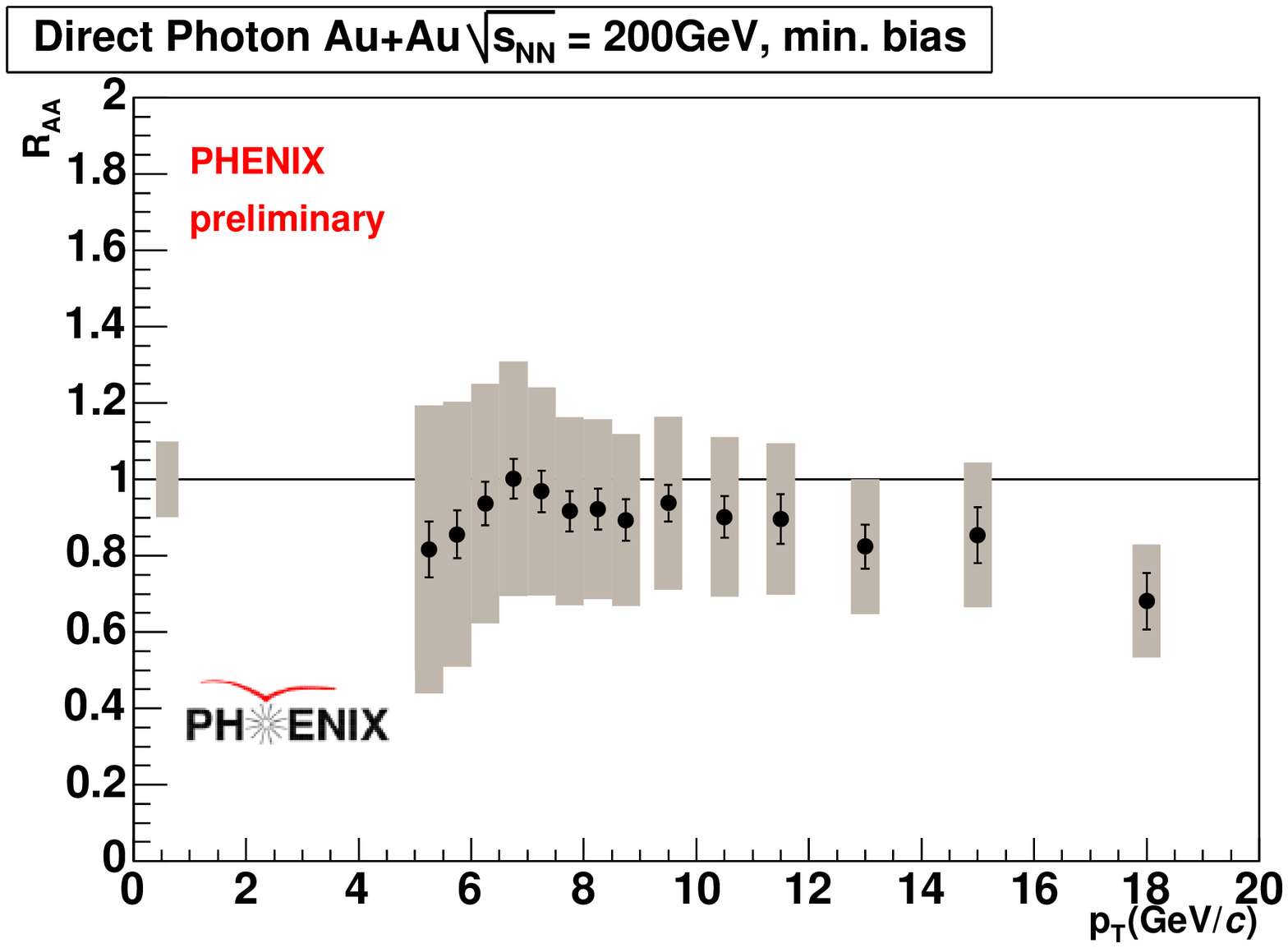}
\end{minipage}
\hspace{5mm}
\begin{minipage}{60mm}
\includegraphics*[width=6.0cm]{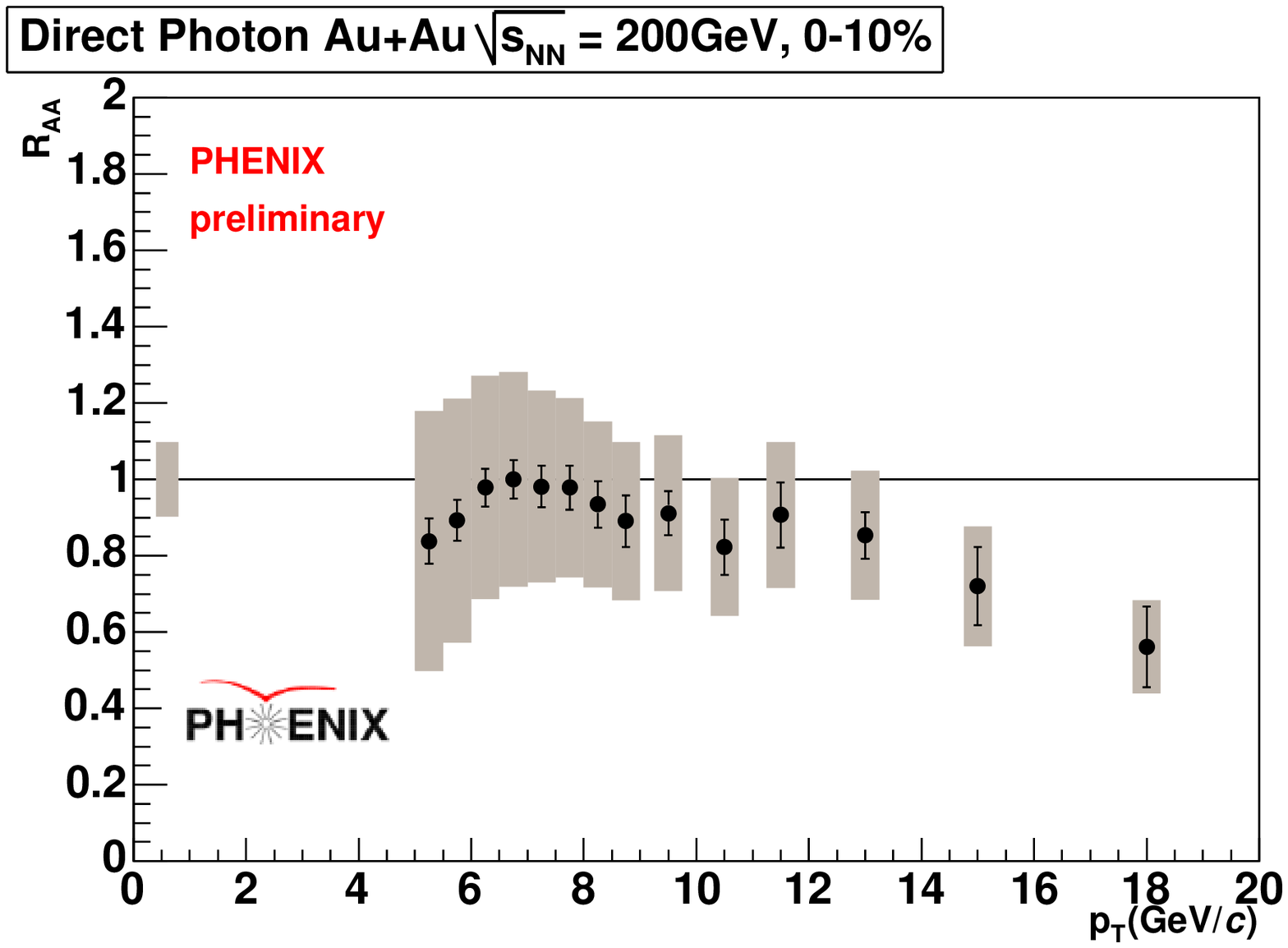}
\end{minipage}
\caption[]{Direct photon $R_{AA}$ in (a) minimum bias (left) and (b) 0-10\,\% (right) Au+Au collisions at $\sqrt{s_{NN}}$=200\,GeV.}
\label{figRAAphotons}
\end{figure}

Direct photon $p_T$ spectra have been measured in Au+Au collisions at
$\sqrt{s_{NN}}$=200\,GeV in RHIC Year-4 run as shown in
Fig.~\ref{figphotonspectra}~\cite{ref8}. The ones in p+p collisions at
the same energy are also shown. The p+p data is used as a baseline
for quantifying the medium effect in Au+Au collisions.
Since the $p_T$ binning is different between Au+Au and p+p results,
the p+p data is fitted with a power-law function to interpolate to the
$p_T$ of the Au+Au data. The fit describes the data very well
within $\sim$5\,\%~\cite{ref8}.
Figure~\ref{figRAAphotons} shows the nuclear modification factor ($R_{AA}$)
of direct photons in minimum bias and 0-10\,\% central Au+Au collisions.
The $R_{AA}$ is defined as:
\[R_{AA} = \frac{(1/N_{evt}) dN/dydp_{T}}{T_{AB}(b) d\sigma^{pp}/dydp_{T}} \]
where $\sigma^{pp}$ is the production cross section in p+p collisions,
and $T_{AB}$ is the nuclear thickness function calculated from the
Glauber model.
With Year-2 run statistics, we were able to reach only up to the region where
the $R_{AA}$ is consistent with unity, and thus concluded that direct photons
are unmodified by the medium~\cite{ref2}. The latest data shows a trend of
decreasing at high $p_T$ ($p_T>$14\,GeV/$c$).

The NLO pQCD calculation predicts that the fragment photons contributes
$\sim$30\,\% to the total at $p_T$$\sim$14\,GeV/$c$. An energy loss of
quarks may result in a suppression of
the yield of the fragment photons. However, the calculation suggests that
the contribution of the fragment photons decreases
as $p_T$ increases, which is contradictory to the observation.

\begin{figure}[htb]
\centering
\begin{minipage}{65mm}
\includegraphics*[width=6.0cm]{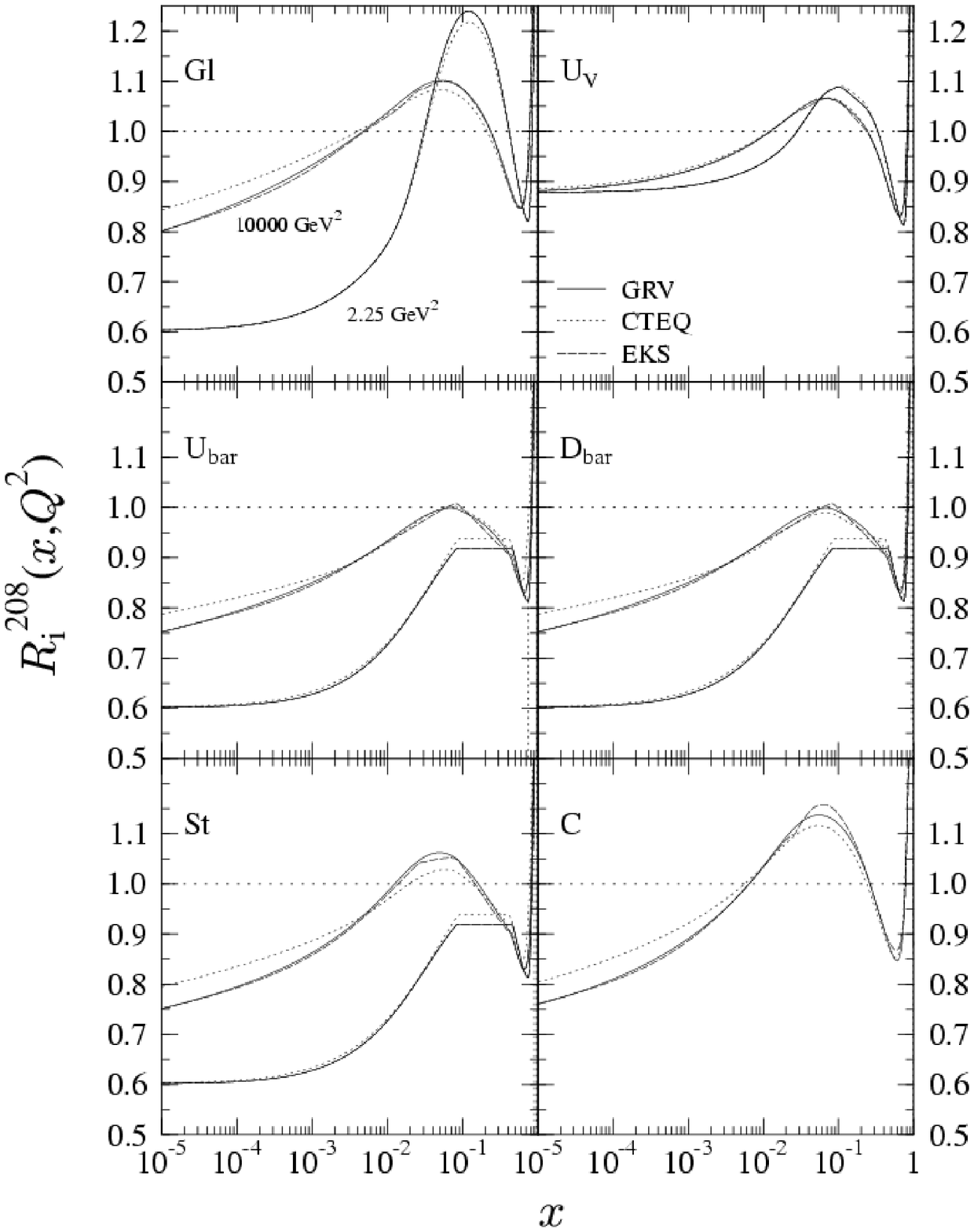}
\vspace{-2mm}
\end{minipage}
\hspace{1mm}
\begin{minipage}{60mm}
\vspace{-1mm}
\includegraphics*[width=5.8cm]{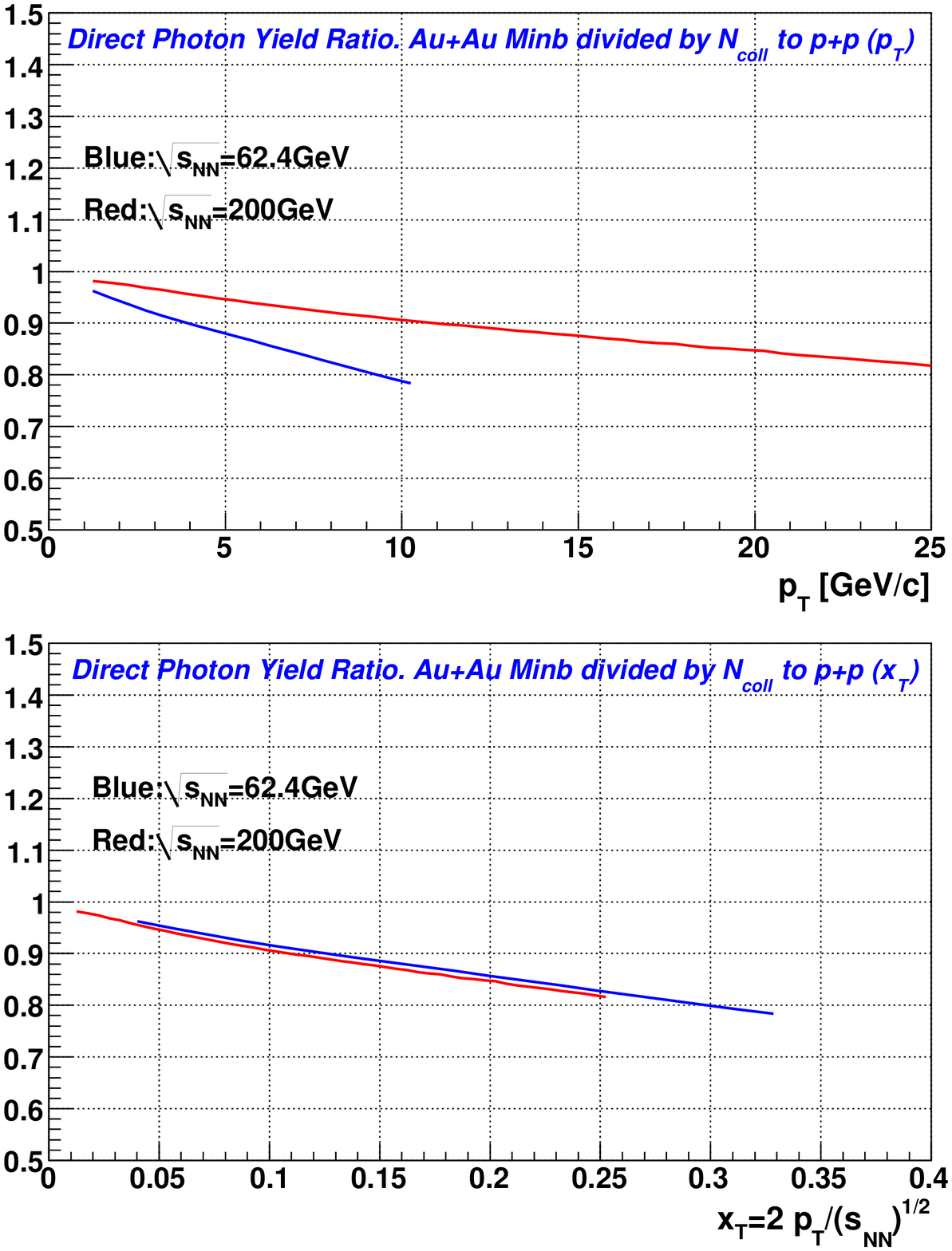}
\vspace{-1mm}
\end{minipage}
\caption[]{(a) Parton distribution functions (left) and $R_{AA}$ of direct photons at $\sqrt{s_{NN}}$=200\,GeV (red) and 62.4\,GeV (blue) as a function of (b) $p_T$ and (c) $x_T$ expected from an isospin effect.}
\label{figStruct}
\end{figure}

Here, we propose simple models to understand the result. The major
contribution to the direct photon production in the $p_T$ range of the
interest is from the Compton scattering process ($qg \rightarrow q \gamma$),
therefore, we can assume that the yield is naively described as:
\[ Yield (x_T, Q^2) = F_{2p}(x_T) \times g_p(x_T) \times \sigma^{dir.\gamma}(x_T, Q^2) \]
where $x_T=2p_T/\sqrt{s}$, $F_{2p}(x_T)$ is the quark parton distribution
function (PDF), and $g_{p}(x_T)$ is the gluon PDF. The $R_{AA}$ can be
written as:
\[ R_{AA} = \frac{d^2 {\sigma_{\gamma}}^{AA}/d{p_T}^2 dy}{AA d^2 {\sigma_{\gamma}}^{pp}/d{p_T}^2 dy} \approx \left(\frac{F_{2A}(x_T)}{A F_{2p}(x_T)} \times \frac{g_{A}(x_T)}{A g_{p}(x_T)} \right) \]
The PDFs are shown in Fig.~\ref{figStruct}(a)\cite{ref9}. 
The decrease of the yield in Au+Au starts at $\sim$12\,GeV/$c$ ($x_T$=0.12)
and drops by $\sim$30\,\% at 18\,GeV/$c$ ($x_T$=0.18).
The PDF does not change by 30\,\% between the two $x_T$ regions.

There is a theoretical calculation that tried to explain the suppression
at high $p_T$ by combining a structure function and
an isospin effect~\cite{ref10}.
The isospin effect is an effect caused by the difference of the quark
charge contents in neutrons and protons. The photon production cross-section
is proportional to $\alpha \alpha_s \Sigma e_q^2$. Therefore, the yield of
photons will be different between p+p, p+n and n+n collisions~\cite{ref5}.
A gold ion consists of 79 protons and 118 neutrons. We can calculate
the hard scattering cross-section for minimum bias Au+Au collisions by
weighting those for p+p ($\sigma^{pp}$), p+n ($\sigma^{pn}$) and
n+n ($\sigma^{nn}$) as follows:
\[\frac{\sigma_{AA}}{<N_{coll}>} = \frac{1}{A^2} \times (Z^2 \sigma_{pp} + 2Z(A-Z) \sigma_{pn} + (A-Z)^2 \sigma_{nn}) \]
where $<N_{coll}>$ is the mean number of binary nucleon-nucleon collisions.
The $R_{AA}$ expected from the isospin effect can be computed as:
\[R_{AA} = \frac{\sigma_{AA}}{<N_{coll}>\sigma_{pp}} \]
The $R_{AA}$ expected from this effect at $\sqrt{s_{NN}}$=200\,GeV is shown
in red in Fig.~\ref{figStruct}(b) and (c). There is $\sim$15\,\% drop at
18\,GeV/$c$ caused by the effect. If we combine the structure function
effect with the isospin effect, the data could be explained.

On the other hand, we could also say that there is almost no suppression
at high $p_T$ within the uncertainty. The opening angle of two $\gamma$'s
decaying from $\pi^0$ becomes very small, and the $\gamma$'s become
indistinguishable at high $p_T$ (starting at $p_T\sim$12\,GeV/$c$)
due to a limited position
resolution (merging effect).

\section{Direct photon measurement in Au+Au collisions at $\sqrt{s_{NN}}$=62.4\,GeV}
\begin{figure}[htbp]
\centering
\includegraphics*[width=7.5cm]{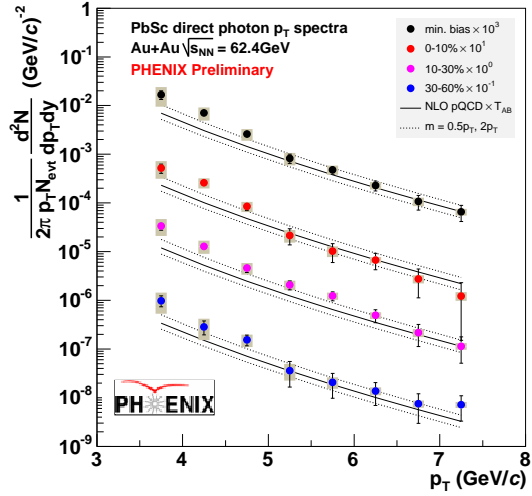}
\vspace{-5mm}
\caption[]{Direct photon $p_T$ spectra in Au+Au collisions at $\sqrt{s_{NN}}$=62.4\,GeV.}
\label{figphoton62}
\end{figure}

\begin{figure}[htbp]
\centering
\begin{tabular}{cc}
\begin{minipage}{70mm}
\includegraphics*[width=5.7cm]{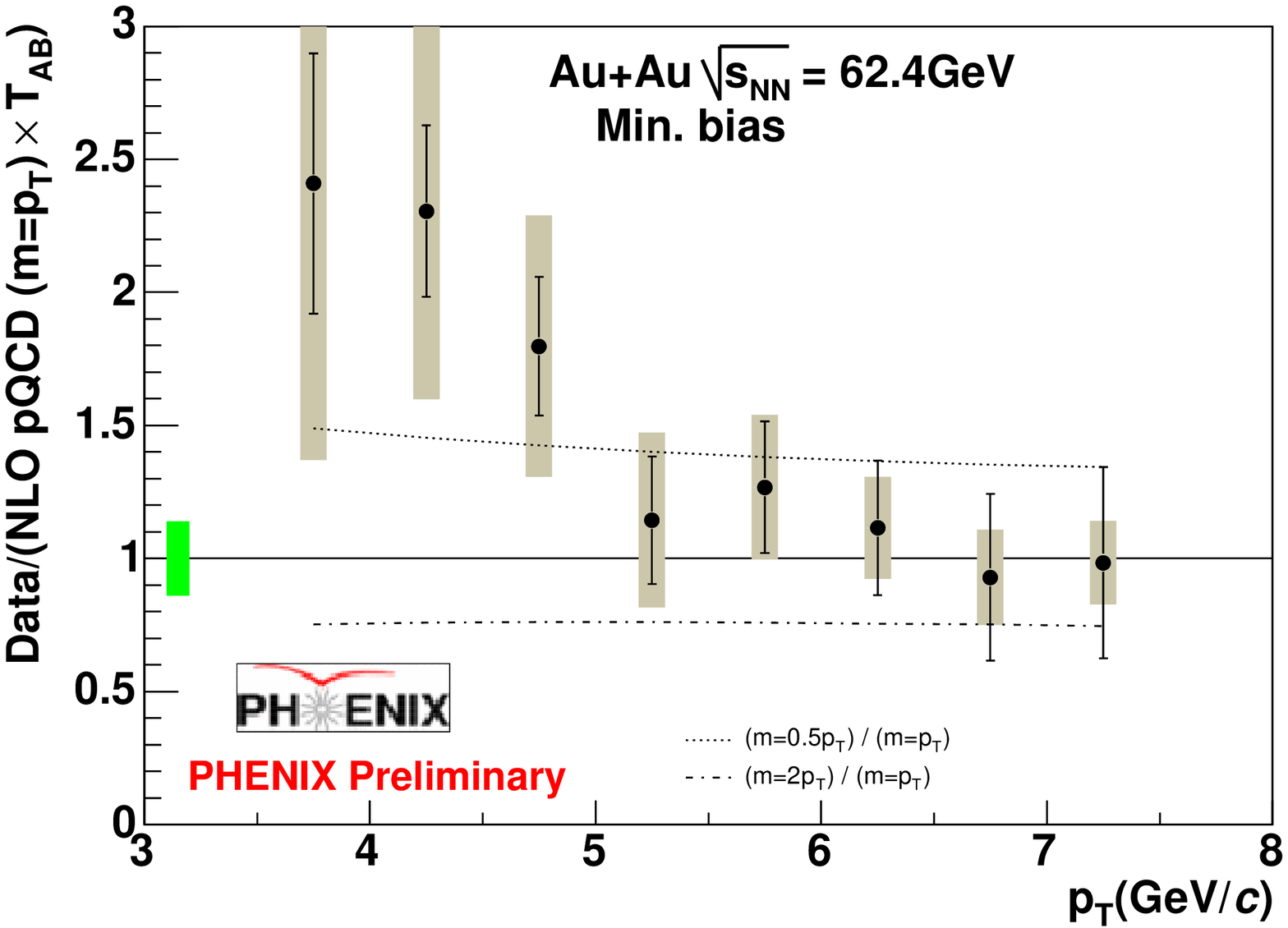}
\end{minipage}
&
\begin{minipage}{70mm}
\includegraphics*[width=5.7cm]{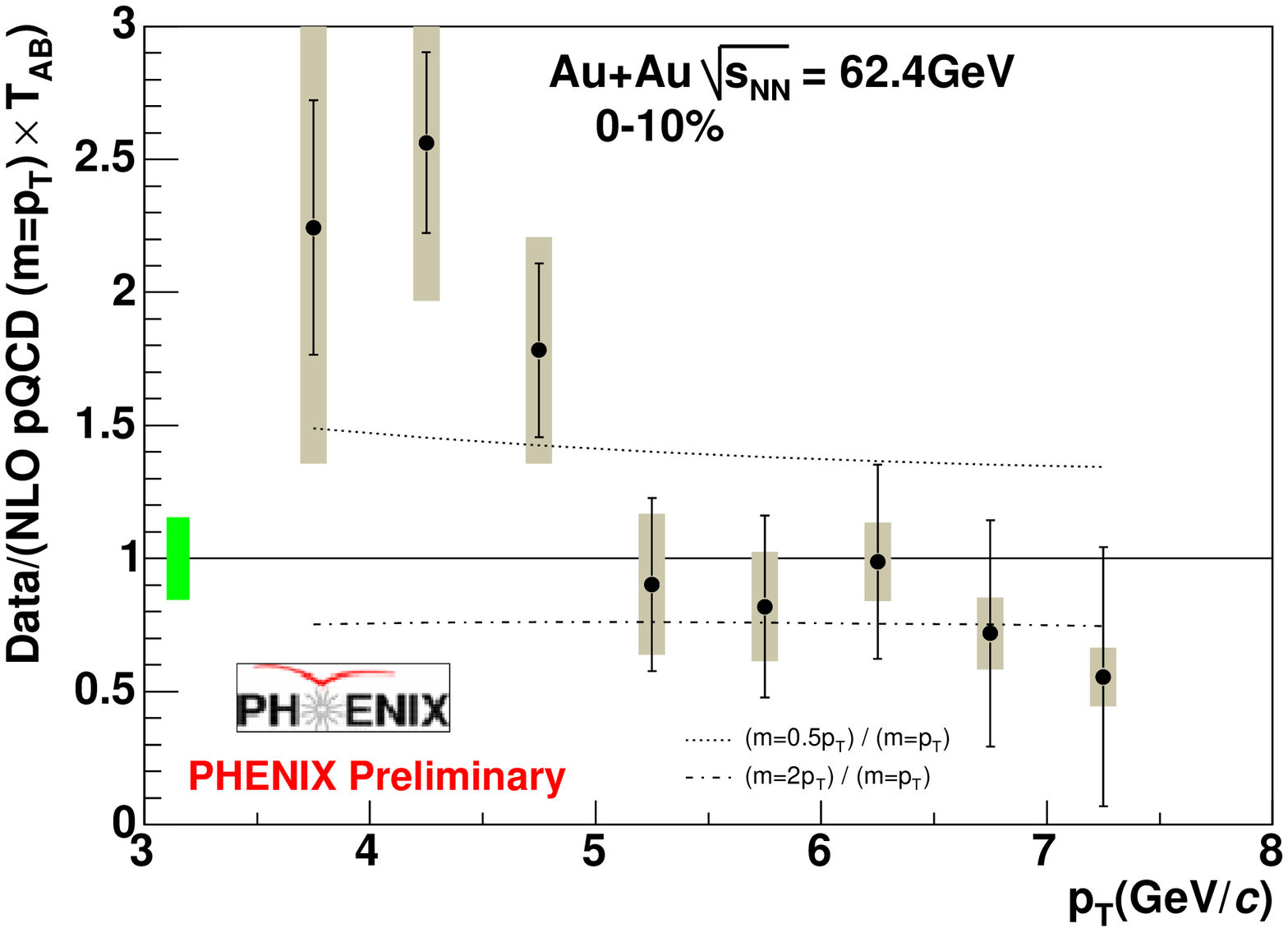}
\end{minipage}
\\
\begin{minipage}{70mm}
\includegraphics*[width=5.7cm]{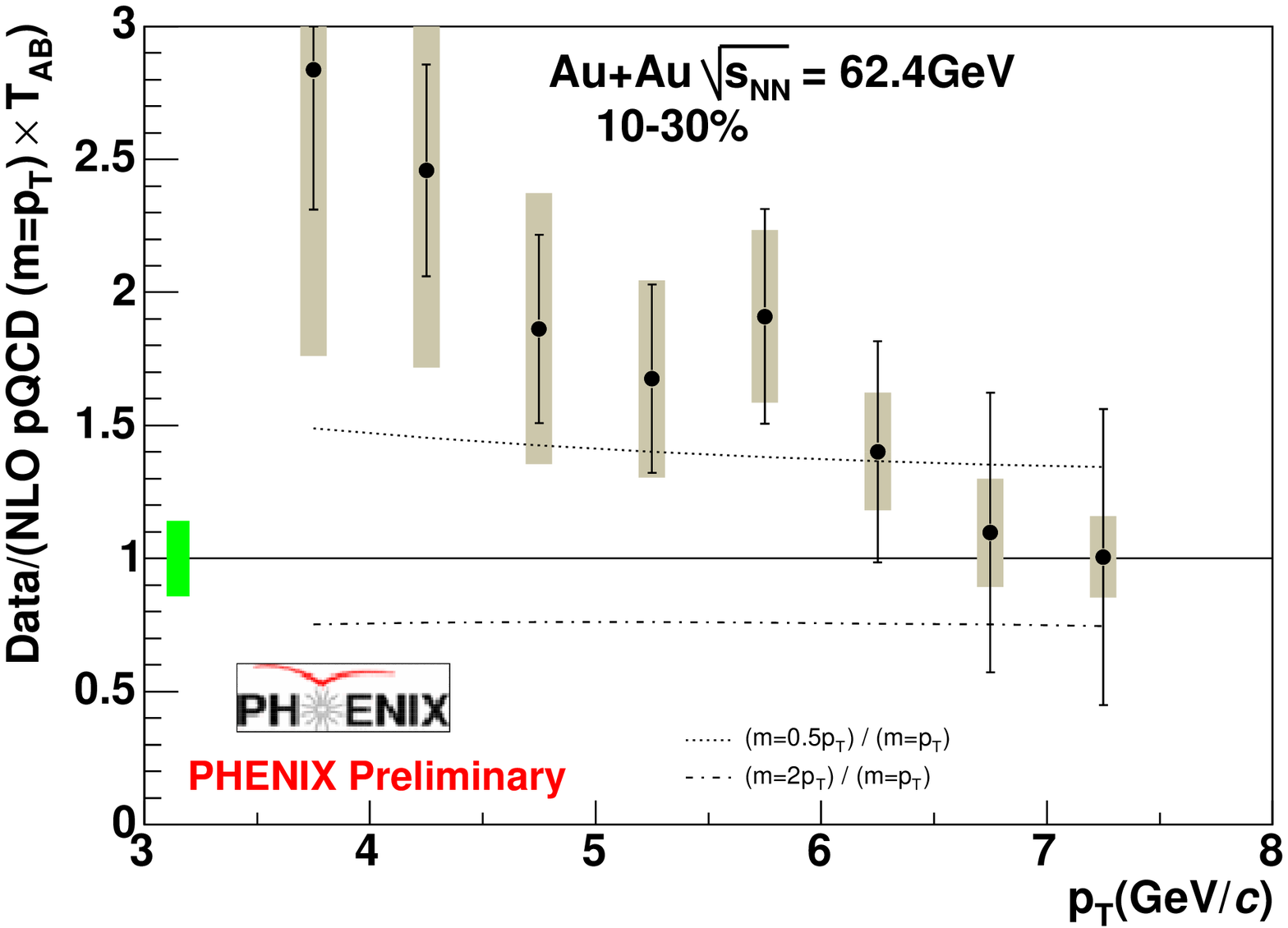}
\end{minipage}
&
\begin{minipage}{70mm}
\includegraphics*[width=5.7cm]{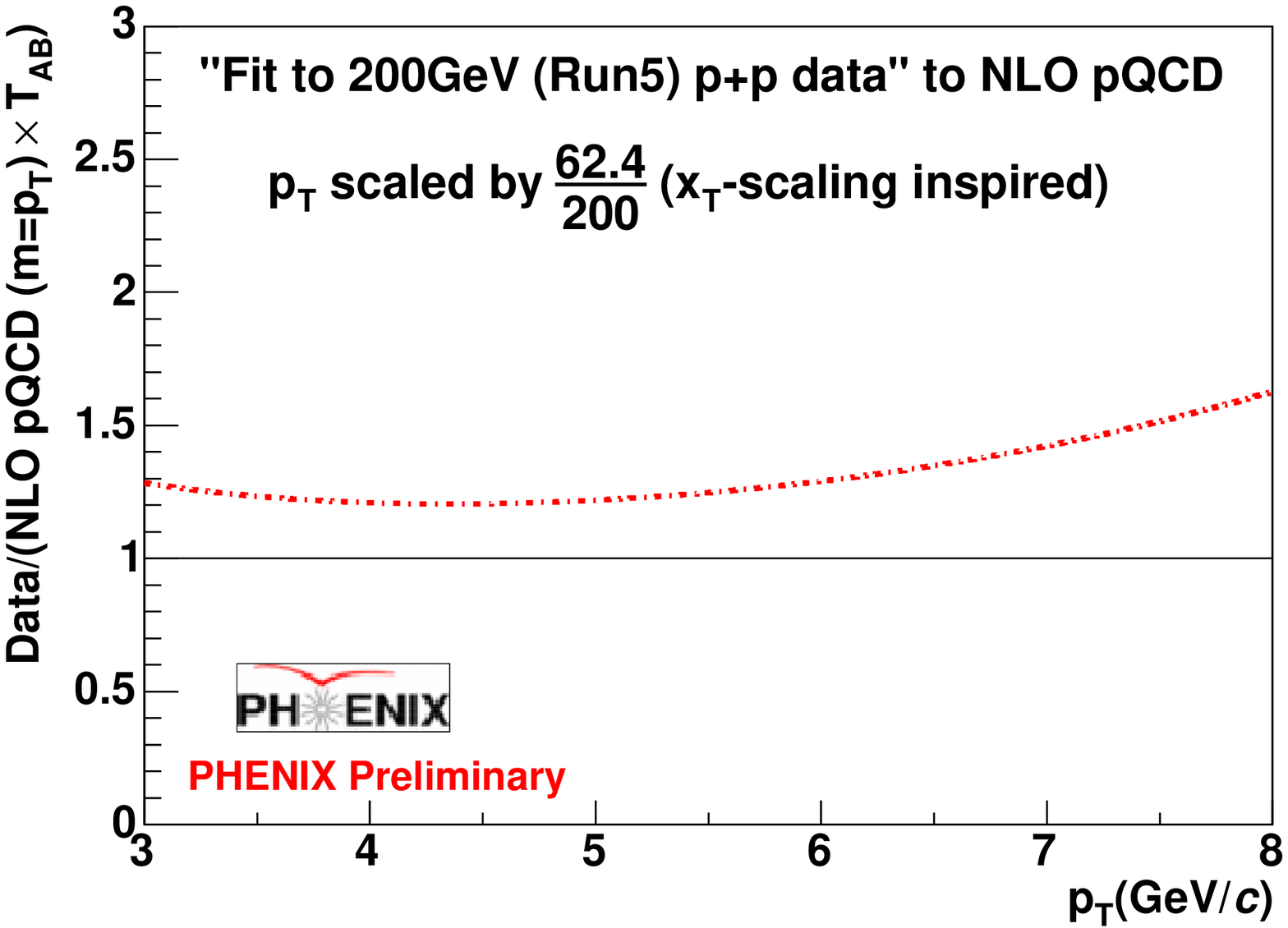}
\end{minipage}
\end{tabular}
\caption[]{Direct photon yield over a NLO pQCD calculation scaled by $T_{AB}$ in Au+Au collisions at $\sqrt{s_{NN}}$=62.4\,GeV. Expected ratio of p+p yield to the calculation is shown in the bottom right.}
\label{figRAA62}
\end{figure}

The $R_{AA}$ expected from the isospin effect at $\sqrt{s_{NN}}$=62.4\,GeV is
calculated in the same way as 200\,GeV, and shown in blue in
Fig.~\ref{figStruct}(b) and (c). The suppression is larger at the same $p_T$
because the effect scales with $x_T$. The calculation suggests that we can
verify the structure function and isospin effect in Au+Au collisions
at $\sqrt{s_{NN}}$=62.4\,GeV without being affected by the merging effect.
Following this argument, direct photons in Au+Au collisions
at $\sqrt{s_{NN}}$=62.4\,GeV have been measured for the first time.
The $p_T$ spectra of the direct photons are shown in Fig.~\ref{figphoton62}.
The NLO pQCD calculation scaled by $T_{AB}$ are overlaid on the data. The data
is qualitatively well described by the calculation.
The ratios of the direct photon yields to the calculation are then
computed and shown in Fig.~\ref{figRAA62}~\footnote{We presented 30-60\,\% centrality in the conference as well, but not shown here because of limited pages.}.
Unfortunately, the yield in p+p collisions at the same energy has not been
measured at any of RHIC experiments. There is a parameterization
of the p+p yield using the measurements at ISR~\cite{ref11}, but the error
is large and systematics is very different. Therefore, we decided not to use
p+p yield for now.

The ratios are essentially consistent with the NLO pQCD expectation within
the experimental and theoretical uncertainty for $p_T>$5\,GeV/$c$. The lower
$p_T$ region may be due to initial multiple scattering of incoming partons
or the jet-photon conversion effect.

It is found that the p+p yield is underestimated by the NLO pQCD calculation
at $\sqrt{s}$=200\,GeV~\cite{ref8}, and it may well happen at 62.4\,GeV.
As an attempt, we took the ratio of the p+p yield to the NLO pQCD calculation
at 200\,GeV, and shifted the horizontal scale by 0.312 (=62.4/200), which
is inspired by an idea of:
\[\frac{\sigma^{pp}}{\sigma^{NLO\ pQCD}} \equiv R,\ \ R_{200GeV}(x_T) = R_{62.4GeV}(x_T)\]
The result is shown in the bottom right of Fig.~\ref{figRAA62}.
If the assumption is correct and the dot-dashed line is considered as
the baseline, we may say that the direct photon yield is suppressed at
high $p_T$ ($p_T>$5\,GeV/$c$,
corresponding to 16\,GeV/$c$ at $\sqrt{s_{NN}}$=200\,GeV), also at
$\sqrt{s_{NN}}$=62.4\,GeV. The $p_T$ region of the interest is immune
from the merging effect, and thus it is considered to be the hint
of physics process such as structure function or isospin effect.
The result is also predicted by a calculation~\cite{ref12}.
Combining the current data with the ones from future d+A and p+A runs would
disentangle these two effects.

\section{Conclusions}\label{concl}
Results on direct photon measurement in Au+Au collisions at
$\sqrt{s_{NN}}$=200\,GeV are discussed from the point of view of
structure function and isospin effect.
The first measurement of
direct photons at $\sqrt{s_{NN}}$=62.4\,GeV in the same collisional
system suggested that these effects are existing and would
manifest at $p_T>$16\,GeV/$c$ at $\sqrt{s_{NN}}$=200\,GeV.



\end{document}